\documentclass[12pt,noshowpacs,nofootinbib,notitlepage,amsmath]{revtex4-2}
\usepackage{setspace}
\allowdisplaybreaks
\usepackage{graphicx,color}
\usepackage[colorlinks=true,citecolor=blue,linkcolor=blue,urlcolor=blue]{hyperref}
\usepackage[charter]{mathdesign}
\usepackage{multirow}
\usepackage{enumerate}
\usepackage{array}
\usepackage{booktabs}

\DeclareSymbolFontAlphabet{\mathcal}{symbols}
\DeclareSymbolFont{symbols}{OMS}{xmdcmsy}{m}{n}
\DeclareSymbolFont{largesymbols}{OMX}{cmex}{m}{n}
\SetSymbolFont{symbols}{bold}{OMS}{xmdcmsy}{b}{n}

\begin{document}  
\title{\color{blue}\Large Photon-photon scattering from a UV-complete gravity QFT}

\author{Bob Holdom}
\email{bob.holdom@utoronto.ca}
\affiliation{Department of Physics, University of Toronto, Toronto, Ontario, Canada  M5S 1A7}
\begin{abstract}
Quantum quadratic gravity (QQG) produces a tree-level differential cross section for $\gamma\gamma\to\gamma\gamma$ that is well-behaved at all energies. From this we can study how the corrections to low energy scattering amplitudes are related to the UV physics, in particular to the exchange of the massive graviparticles. An effective forward scattering amplitude is obtained by separating out the effects of the $t$-channel graviton pole. This is possible due to the UV-completeness, and even though the Froissart bound is not satisfied. We then consider photon-photon scattering to two graviparticles and a further imaginary contribution to the $\gamma\gamma\to\gamma\gamma$ forward scattering amplitude. Unitarity without positivity is a key property of QQG and it impacts all our results.
\end{abstract}

\maketitle 

\section{Introduction}

Quantum quadratic gravity (QQG) is a possible UV complete QFT of gravity that was shown early to be renormalizable \cite{stelle}. QQG describes a massless graviton, a massive spin-2 ghost and a massive scalar, and we shall refer to these particles collectively as graviparticles. The theory may be defined in terms of the ghost mass $m_G$ and the graviscalar mass $m_S$,
\begin{align}
S&=-\frac{1}{16\pi G}\int d^4x \sqrt{-g}\left(R+ \frac{R_{\mu\nu}R^{\mu\nu}-\frac{1}{3}R^2}{m_G^2}-\frac{R^2}{6m_S^2} \right).
\label{e17}\end{align}
These masses are naturally of order the Planck mass $m_{\rm Pl}$. The quantity $G m_G^2$ is a dimensionless coupling, and it is an asymptotically free coupling, even when including all matter contributions to the running \cite{fradkin1,fradkin2,Salvio:2014soa}. The graviton exists in both the IR and UV limits of the theory. Both of these limits are well behaved due to the massive graviparticles that alter and improve the UV behavior of the theory above their mass scale. Unlike QCD, where the quarks and gluons are appropriate degrees of freedom only for the UV limit, the perturbative degrees of freedom of QQG are sufficient to describe both the IR and UV limits.

The perturbative theory satisfies the usual requirements of locality, unitarity and analyticity. What the theory lacks is positivity. Both the lack of positivity and the renormalizability of the theory originates in the same place, namely a relative minus sign between the massive spin-2 propagator and the graviton propagator. The lack of positivity has two further implications. One is that the dressed ghost propagator is such that the ghost decays backward in time. The other is that certain exclusive cross sections involving the ghost are negative. These two apparent problems cancel each other. The former shows that the ghost is not a true asymptotic state, and so the particular cross sections that are negative are not physically meaningful on their own. On the other hand these minus signs end up ensuring good high energy behavior of suitably \textit{inclusive} cross sections \cite{Holdom:2021hlo}. This is how the theory ensures good high energy behavior of physically relevant cross sections, as might be expected in a theory that is unitarity, renormalizable and perturbatively predictive. Cancellations that occur at the level of cross sections is not a new concept in QFT; it is well known in the treatment of infrared divergences.

In \cite{Holdom:2021hlo} we studied the ultra-Planckian scattering of graviparticles and we developed a parton-shower-like description. The hard scattering subprocesses are described by inclusive differential cross-sections and we found these to be well-behaved. In QCD the cross sections for jet production are similarly obtained from suitably inclusive cross sections among the quarks and gluons, the partons in this case, even though the partons are not the asymptotic states. In QQG the partons are the graviparticles, and the spin-2 ghost in particular is not an asymptotic state. The relevance of inclusive partonic cross sections is due to an effective duality between all possible partonic states and all possible physical states. We also learn from QCD that partonic cross sections apply even though the actual hard subprocess scatters the \textit{off-shell} partons participating in the initial state or final state parton showers. In QQG it is with this off-shell description that we see cancellations at the amplitude level that are occurring because of differing signs of off-shell graviparticle propagators. This is another way of seeing how good high energy behavior can follow, in a way similar to how renormalizability follows. The differing signs and cancellations among exclusive partonic cross sections is just another manifestation of this.

The topic of this paper is more straightforward. We are interested in the corrections to the low energy theory in the form of powers of $E/m_{\rm Pl}$ that result from the UV completion, in particular from the tree-level exchange of massive graviparticles. These corrections should provide an imprint of the good high energy behavior that this theory enjoys. But something unusual happens when this issue is explored in the context of graviton-graviton scattering. In the particular case of $gg\to gg$ tree-level amplitudes, QQG produces results that are identical to those produced by GR \cite{Dona:2015tra}. Thus in this case there are no $E/m_{\rm Pl}$ corrections, although presumably they would show up at one-loop level. We will thus turn to photon-photon scattering to explore the relation between IR corrections and the UV theory. 

It has become popular to establish constraints on the IR corrections even when the UV theory is not known, e.g.~\cite{Adams:2006sv,Cheung:2014ega,Camanho:2014apa,Alberte:2020bdz,Caron-Huot:2021rmr}. These constraints follow from dispersion relations along with the analyticity and unitarity of the UV theory. Positivity is also crucial for these results, and this is either just assumed implicitly, or it is assumed to be a consequence of unitarity. Perturbative QQG is an explicit counter-example that enjoys analyticity and unitarity but not positivity. Thus it is of interest to explore our results in the context of these standard methods. In addition we are able to more easily address the issue of the $t$-channel pole in the forward scattering amplitude, due to the massless graviton, since we are working in the context of a UV complete theory.

The following propagator describes the propagation of the three graviparticles (graviton, ghost $G$ and graviscalar $S$),
\begin{align}
G_{\mu\nu\rho\sigma}&=i16\pi G\left(-\frac{2m_G^2}{q^2(q^2-m_G^2)}P^2_{\mu\nu\rho\sigma}+\frac{m_S^2}{q^2(q^2-m_S^2)}P^0_{\mu\nu\rho\sigma}\right),\label{e11}\\\nonumber
&P^2_{\mu\nu\rho\sigma}=\frac{1}{2}(\theta_{\mu\rho}\theta_{\nu\sigma}+\theta_{\mu\sigma}\theta_{\nu\rho})-\frac{1}{3}\theta_{\mu\nu}\theta_{\rho\sigma},\\\nonumber
&P^0_{\mu\nu\rho\sigma}=\frac{1}{3}\theta_{\mu\nu}\theta_{\rho\sigma}
\quad\theta_{\mu\nu}=\eta_{\mu\nu}-\frac{q_\mu q_\nu}{q^2}
.\end{align}
For $\gamma\gamma\to\gamma\gamma$ there are three diagrams with graviparticles exchanged in the $s$, $t$ and $u$ channels. We display the required graviparticle-photon-photon vertex in the appendix. In addition to $\gamma\gamma\to\gamma\gamma$ we shall also consider $\gamma\gamma\to{\cal G}^S{\cal G}^S$ where ${\cal G}^S$ denotes any of the graviparticles, $g$, $G$ or $S$. Here there are two diagrams with a photon exchanged in the $t$ and $u$ channels, one with a graviparticle in the s-channel, and one contact diagram. The additional vertices for this process are displayed in the appendix, at least for vertices involving photons.

It is worth stressing that the usual Feynman $i\varepsilon$ prescription applies at all poles of the graviparticle propagator, and thus only the overall sign of the ghost pole is unusual. This is key to the renormalizability of the theory \cite{stelle}. It is also associated with perturbative unitarity, that is with the standard perturbative derivation of the optical theorem. The optical theorem in QQG reflects the lack of positivity by having new minus signs in some corresponding terms on either side of the relation \cite{Holdom:2021hlo}. Up to such minus signs, the analytic properties of Feynman diagrams is standard. We previously mentioned the \textit{dressed} ghost propagator that is obtained via a resummation, and this yields nonstandard analytic properties corresponding to backward in time decay. But the perturbative expansion utilizes the \textit{bare} propagator in (\ref{e11}), and the perturbative results order by order will have standard analytic properties.

\section{Photon-photon scattering to two photons}
The differential cross-section for $\gamma\gamma\to \gamma\gamma$ is
\begin{align}
\frac{d\sigma}{d\Omega}=\frac{1}{64\pi^2 s}\frac{1}{4}|{\cal M}|^2,
\end{align}
where sums over polarizations are implicitly present in $|{\cal M}|^2$ and the $1/4$ is due to the average over initial polarizations. We shall present various results for $|{\cal M}|^2$ in terms of Mandelstam variables, which can be transformed back to CoM variables $E$ and $\theta$ if desired. Our full tree-level gravity-induced result, valid at all energies, is
\begin{align}
|{\cal M}|^2&=32\pi^2G^2m_G^4\frac{F_1+F_2m_G^{2}+F_3m_G^{4}+F_4 m_G^{6}+F_5 m_G^{8}}{u^{2} t^{2} s^{2} (s-m_G^{2} )^{2} (t-m_G^{2} )^{2} (u-m_G^{2} )^{2}},\label{e16}\\
&F_1=\frac{1}{32} v^{6}-\frac{1}{2} w^{2} v^{3}+3 w^{4},\\
&F_2=\frac{1}{2}w v (v^{3}-8 w^{2}),\\
&F_3=-\frac{1}{8}v^{2} (v^{3}-12 w^{2}),\\
&F_4=-w (v^{3}+12 w^{2}),\\
&F_5=\frac{1}{8}v (v^{3}+32 w^{2})=s^8+t^8+u^8,
\end{align}
where $v=s^2+t^2+u^2$ and $w=s t u$. The ghost mass $m_G$ appears, but not the graviscalar mass $m_S$, even though the full graviparticle propagator has been used. We will return to this point in Section \ref{s4} where $S$ appears in the final state.

The low energy limit ($m_G\to \infty$) is
\begin{align}
|{\cal M}|^2&=32\pi^2G^2\frac{s^{8}+t^{8}+u^{8}}{s^{2} t^{2} u^{2}}
.\label{e1}\end{align}
When written in the CoM frame in terms of $E=\sqrt{s}/2$ and the scattering angle $\theta$, this is
\begin{align}
|{\cal M}|^2&=(8\pi G)^2E^4\frac{\cos(\theta)^{8}+28 \cos(\theta)^{6}+70 \cos(\theta)^{4}+28 \cos(\theta)^{2}+129}{\sin(\theta)^{4}}.
\end{align}
This agrees with eq.~(15) of \cite{Brodin:2006wa}. At low enough energy this result will dominate the electron loop contribution, since that falls faster in the infrared like $|{\cal M}|^2\sim\alpha^4 E^8/m_e^8$. The latter also doesn't have the $1/\sin(\theta)^4$ enhancement.

The high energy limit ($m_G\to 0$) is
\begin{align}
|{\cal M}|^2&=32\pi^2G^2m_G^{4}\frac{v^6/32-w^2 v^3/2+3 w^4}{s^{4} t^{4} u^{4}}.
\label{e2}\end{align}
This remains bounded in the high energy limit when $s,t,u\to\infty$. It yields a differential cross section $d\sigma/d\Omega$ that falls like $1/E^2$ as $E\to\infty$, as might be hoped for in a UV complete QFT for gravity. (Similar good high energy behavior for scalar-scalar scattering in QQG was reported in \cite{Abe:2017abx}.) Compared to the low energy result, there is an even stronger $1/(t^4 u^4)\sim 1/\sin(\theta)^8$ enhancement in the forward and backward scattering limits. We see from the full result (\ref{e16}) how $1/(t^2 u^2)$ behavior goes to $1/(t^4 u^4)$ behavior when $t,u\to\infty$.

Since we expect that the gravitational coupling $Gm_G^2$ is of order one, (\ref{e2}) can easily dominate over the high energy contributions from charged matter loops that are suppressed by $\alpha^4$. The one-loop corrections to our tree-level result will cause $Gm_G^2$ to effectively run and become weaker in the deep UV. Any matter particle or graviparticle in the loop contributes to the weakening effect.\footnote{In \cite{Alberte:2020bdz} it is argued that a particular contribution to photon-photon scattering, the electron-loop vertex correction to the graviton exchange diagram, produces some problem at high energy. We do not agree with that assessment.} Thus in the deep UV it could be that the coupling $G m^2_G$ becomes weaker than some other asymptotically-free matter coupling. But at least in some energy range, gravity dominates ultra-Planckian scattering.

Let us turn to a study of the amplitudes. We will denote the polarization vector of each photon by t when it is in the scattering plane, and by u when it is orthogonal. For the various combinations of polarizations there are four amplitudes,
\begin{align}
\label{e4}&\mbox{tttt or uuuu:}\quad g(s,t,u)\\
\label{e6}&\mbox{tutu or utut:}\quad h(t;s,u)\\
\label{e5}&\mbox{ttuu or uutt:}\quad h(s;t,u)\\
\label{e7}&\mbox{tuut or uttu:}\quad h(u;s,t)
\end{align}
The tttt or uuuu amplitude involves four identical particles and thus $g(s,t,u)$ is $s$-$t$-$u$ symmetric. The other three amplitudes are respectively $s$-$u$, $t$-$u$, $s$-$t$ symmetric. The two functions are, with a factor of $8\pi G$ implicit,
\begin{align}
&g(s,t,u)\equiv\frac{\frac{1}{4}(v^3-12w^2)m_G^2+2 v w m_G^{4}+\frac{1}{2}v^2 m_G^{6}}{4 s t u (s-m_G^{2} ) (t-m_G^{2} ) (u-m_G^{2} )},\\
&h(t;s,u)\equiv\nonumber\\
&\frac{(-2 s^{4}-5 s^{3} u -8 s^{2} u^{2}-5 s \,u^{3}-2 u^{4}) m_G^{2}+(4 s^{2} u +4 s \,u^{2}) m_G^{4}+(4 s^{2}+6 s u +4 u^{2}) m_G^{6}}{4 t(s-m_G^{2} ) (t-m_G^{2} ) (u-m_G^{2} )}.
\end{align}

Expanding in powers of $1/m_G^2$ gives
\begin{align}
g(s,t,u)&=\frac{(s^{2}+t^{2}+u^{2})^2}{8 s t u}-\frac{s^{2}+t^{2}+u^{2}}{2m_G^{2}}+\frac{3 s t u}{4 m_G^{4}}+\cdots,\\
h(t;s,u)&=-\frac{2 s^{2}+3 s u +2 u^{2}}{2 t}+\frac{s u}{m_G^{2}}-\frac{t (2 s^{2}+s u +2 u^{2})}{4 m_G^{4}}+\cdots
.\end{align}
The first terms are present in GR, and we can compare and expand these GR terms of all four amplitudes (\ref{e4}-\ref{e7}),
\begin{align}
\frac{(s^{2}+t^{2}+u^{2})^2}{8 s t u}&=-\frac{s^{2}}{2} t^{-1}-\frac{1}{2} s -t +\frac{t^3}{2su},\label{e12}\\
-\frac{2 s^{2}+3 s u +2 u^{2}}{2 t}&=-\frac{s^{2}}{2} t^{-1}-\frac{1}{2} s -t\label{e13},\\
-\frac{2 t^{2}+3 t u +2 u^{2}}{2 s}&=-s -\frac{1}{2} t - \frac{1}{2s} t^{2},\\
-\frac{2 s^{2}+3 s t +2 t^{2}}{2 u}&=s +\frac{1}{2} t +\frac{1}{2s} t^{2}+\frac{t^3}{2su}.
\end{align}
If we wish to take the $t\to 0$ limit naively then these results show that this can only be done for (\ref{e5}) and (\ref{e7}). Taking this limit for (\ref{e5}) allows us to have a first look at $s/m_G^2$ corrections,
\begin{align}
-\frac{s}{2}\frac{(s^{2}-2 m_G^{4})}{s^{2}-m_G^{4}}=-s - \frac{1}{2m_G^{4}} s^{3}-\frac{1}{2m_G^{8}} s^{5}-\frac{1}{2m_G^{12}} s^{7}+{\cal O}(s^{9})
.\label{e9}\end{align}
(\ref{e7}) is the same but without the overall minus sign. The fact that both signs occur is related to the fact that the $t\to0$ limits of (\ref{e5}) and (\ref{e7}) are not \textit{elastic} forward scattering amplitudes, which thus prevents their use in the optical theorem. ((\ref{e7}) represents elastic scattering in the backward limit, but this gives the pole $-\frac{1}{2}s^2/u$.)

\section{Forward scattering amplitudes}
Thus we must focus on (\ref{e4}) and (\ref{e6}), and the first thing we can do is to look at their large $s$ behavior for fixed $t$. The leading term at large $s$ for both amplitudes is
\begin{align}
\frac{m_G^{2} s^{2}}{2 t (t-m_G^{2})}=-\frac{s^{2}}{2} t^{-1}-\frac{1}{2} \frac{s^{2}}{m_G^{2}}-\frac{1}{2} \frac{s^{2}}{m_G^{4}} t -\frac{1}{2} \frac{s^{2}}{m_G^{6}} t^{2}+{\cal O}(t^{3})
,\label{e15}\end{align}
which we have then expanded in $t$. We continue to leave the $8\pi G$ factor implicit. Thus along with the $-\frac{1}{2}s^2/t$ pole there is another $s^2$ term that survives as $t\to0$, coming from the UV completion. This $-\frac{1}{2}s^2/m_G^2$ contribution to the forward scattering amplitude means that the Froissart bound is not satisfied. This is despite the locality, analyticity and unitarity of a theory that produces bounded amplitudes when $s,t,u\to\infty$. Although the Froissart bound is often assumed to apply to gravity theories, we see from (\ref{e15}) that it is violated for the simple reason that the UV completion generates a massive $t$-channel pole that accompanies the massless $t$-channel pole.

We may define the full UV contribution to the forward scattering amplitude as
\begin{align}
{\cal M}^{UV}(s)=\left.\left({\cal M}(s,t)-{\cal M}^{GR}(s,t)\right)\right|_{t=0}
.\end{align}
Using either (\ref{e4}) or (\ref{e6}) for ${\cal M}(s,t)$, with the associated (\ref{e12}) or (\ref{e13}) for ${\cal M}^{GR}(s,t)$, we find
\begin{align}
{\cal M}^{UV}(s)&=-\frac{s^2}{2m_G^2}\frac{(s^{2}-2 m_G^{4})}{s^{2}-m_G^{4}}\label{e10}\\&=-\frac{1}{m_G^2}s^2 - \frac{1}{2m_G^{6}} s^{4}-\frac{1}{2m_G^{10}} s^{6}-\frac{1}{2m_G^{14}} s^{8}+{\cal O}(s^{10})
.\end{align}
Thus the UV completion contributes to the forward scattering amplitude in such a way that the low energy expansion parameters are negative definite.

Before discussing the meaning of this, it is interesting to put it into the context of standard approaches that are used when, unlike our case, the UV completion is not known. If ${\cal M}(s,t=0)$ is an elastic forward scattering amplitude, then positivity constraints on expansion parameters are usually found by considering the behavior of ${\cal A}_n(s)={\cal M}(s,t=0)/s^n$ on the complex $s$ plane, with $n=-3$ being the typical choice. It is a well known problem for gravity that ${\cal M}(s,t=0)$ is not defined due to the $-\frac{1}{2}s^2/t$ pole, as is the case for (\ref{e4}) and (\ref{e6}). We shall approach this problem by holding $t$ fixed at some small negative value in such a way that $t\to0$ can taken at the end.

Noting that $u=s$ when $s=-t/2$, we introduce
\begin{align}
{\cal A}_{(n,t)}(s)=\frac{{\cal M}(s,t)}{(s+t/2)^n}.
\label{e14}\end{align}
This quantity will help us to separate the IR corrections of interest from the effects of the $-\frac{1}{2}s^2/t$ pole. ${\cal A}_{(n,t)}(s)$ is a rational function of $s$ and so we can make use of an identity that relates the residues of its simple poles on the complex $s$ plane. Since we know the UV completion we will be able to handle effects that arise at asymptotically large values of $|s|$. This may be done by adding the point at $\infty$ to the complex plane to render it compact. Then if $s=p_i$ are the pole locations and ${\rm Res}_{p_i}{\cal A}_{(n,t)}(s)$ is the residue of the $i$th pole we have
\begin{equation}
\sum_i {\rm Res}_{p_i}{\cal A}_{(n,t)}(s)+{\rm Res}_\infty {\cal A}_{(n,t)}(s)=0
.\label{e3}\end{equation}
The residue of a possible pole at $\infty$ may be calculated via ${\rm Res}_\infty {\cal A}_{(n,t)}(s)\equiv -{\rm Res}_0 {\cal A}_{(n,t)}(1/s)/s^2$. With this we have a relation that is true for any $n$ at fixed $t$. Violation of the Froissart bound is of no concern.

\def\arraystretch{1.3}
\begin{table}[h]
\centering
$$\begin{array}{|c||c|c|c|c||c|c|c|c|}
\hline
n & 0 & -t & -t/2 & \infty & -t/2 & \infty & m_G^2 & -m_G^2-t \\\hline\hline
 -2 & 0 & 0 & 0 & 0 & 0 & 0 & \frac{1}{4}m_G^{8} & -\frac{1}{4}m_G^{8} \\\hline
  -1 & 0 & 0 & 0 & 0 & 0 & -\frac{1}{2}m_G^{6} & \frac{1}{4}m_G^{6} & \frac{1}{4}m_G^{6}\\\hline
 0 & 0 & 0 & 0 & 0 & 0 & 0 & \frac{1}{4}m_G^{4} & -\frac{1}{4}m_G^{4} \\\hline
  1 & 0 & 0 & 0 & 0 & 0 & -\frac{1}{2}m_G^{2} & \frac{1}{4}m_G^{2} & \frac{1}{4}m_G^{2} \\\hline
 2 & -2 & 2 & 0 & 0 & 0 & 0 & \frac{1}{4} & -\frac{1}{4} \\\hline
  3 & -4t^{-1} & -4t^{-1} & \frac{15}{2}t^{-1} & \frac{1}{2}t^{-1} & -m_G^{-2} & \frac{1}{2}m_G^{-2} & \frac{1}{4}m_G^{-2} & \frac{1}{4}m_G^{-2} \\\hline
 4 & -8t^{-2} & 8t^{-2} & 0 & 0 & 0 & 0 & \frac{1}{4}m_G^{-4} & -\frac{1}{4}m_G^{-4} \\\hline
  5 & -16t^{-3} & -16t^{-3} & 32t^{-3} & 0 & -\frac{1}{2}m_G^{-6} & 0 & \frac{1}{4}m_G^{-6} & \frac{1}{4}m_G^{-6} \\\hline
 6 & -32t^{-4} & 32t^{-4} & 0 & 0 & 0 & 0 & \frac{1}{4}m_G^{-8} & -\frac{1}{4}m_G^{-8} \\\hline
7 & -64t^{-5} & -64t^{-5} & 128t^{-5} & 0 & -\frac{1}{2}m_G^{-10} & 0 & \frac{1}{4}m_G^{-10} & \frac{1}{4}m_G^{-10} \\\hline
\end{array}
$$\caption{Residues of poles of ${\cal A}_{(n,t)}(s)$ in (\ref{e14}). The pole locations are given at the top. The poles at $-t/2$ and $\infty$ each have two columns since there are two types of contributions to their residues. The left half of the table is due GR and the right half is due to the UV completion.}\label{t1}
\end{table}

We first take ${\cal M}(s,t)$ to be the tttt or uuuu amplitude in (\ref{e4}) with $u=-s-t$. Then in Table \ref{t1} we show how (\ref{e3}) is satisfied for various $n$. The explicit poles in ${\cal M}(s,t)$ are the two at 0 and $-t$ which correspond to the massless $s$ and $u$ poles, and the two at $m_G^2$ and $-m_G^2-t$ which correspond to the massive $s$ and $u$ poles. There is also the pole structure introduced in (\ref{e14}) at $-t/2$, as well as the possible pole at infinity. Table \ref{t1} shows all contributions to the residues of these poles that survive the $t\to0$ limit. There are two types of contributions to the $-t/2$ and $\infty$ residues, the contributions singular for $t\to\infty$ in the third and fourth columns and the nonsingular contributions in the fifth and sixth columns. We see that $n=3$ is the only case where there are residues at both $-t/2$ and $\infty$ simultaneously.

If we instead take ${\cal M}(s,t)$ to be the amplitude in (\ref{e6}), the massless $u$ and $s$ poles do not exist, and the result is that the only singular contributions occur for $n=3$ where there are $-\frac{1}{2}t^{-1}$ and $\frac{1}{2}t^{-1}$ entries in the third and fourth columns. The right half of the table remains the same as in Table \ref{t1}. We may also take ${\cal M}(s,t)$ to be the purely GR results in (\ref{e12}) and (\ref{e13}). Then the results for the left half of Table \ref{t1} are identical to the respective results for (\ref{e4}) and (\ref{e6}), while the right half of the table for (\ref{e12}) and (\ref{e13}) are identically zero. Thus we have a clean separation between the effects that are purely due to GR, the left half of the table, and the effects that are purely due to the UV completion, the right half of the table. All effects singular in $1/t$ are in the left half.

Most importantly, the effects in each half of the table cancel amongst themselves for every $n$ with respect to relation (\ref{e3}). Thus we can now safely take $t\to0$, which means that the pole locations for the fifth and eighth columns are moving to $0$ and $-m_G^2$ respectively. The fifth column gives the coefficients of  a Taylor series expansion in $s$, and the table relates these to the UV physics as reflected by the residues of the massive poles and the pole at $\infty$. In fact our result for ${\cal M}^{UV}(s)$ in (\ref{e10}) reproduces the right half of the table via ${\cal A}_n^{UV}(s)={\cal M}^{UV}(s)/s^n$, with pole locations at 0, $\infty$, $m_G^2$ and $-m_G^2$. This reinforces ${\cal M}^{UV}(s)$ as being the full UV contribution to the forward scattering amplitude.

Not let us discuss the meaning of our result for ${\cal M}^{UV}(s)$. Since the usual Feynman $i\varepsilon$ prescription applies in QQG, as mentioned in the Introduction, the sign of the residue of a pole is opposite in sign to the imaginary part of the pole. In turn from the optical theorem, a particular imaginary contribution to a forward elastic scattering amplitude is related to a particular exclusive cross-section, in this case for the production of the particle giving the pole. From (\ref{e10}), the pole of ${\cal M}^{UV}(s)$ at $s=m_G^2$ has positive residue, and this points to a negative exclusive cross section for the production of a ghost. As discussed in \cite{Holdom:2021hlo}, the perturbative theory produces negative cross-sections for processes with an odd number of ghosts. This is permissible because the ghost does not survive as an asymptotic state; its dressed propagator describes a state that decays backwards in time. Associated with this is a negative decay width, and so the minus signs cancel in the production and decay of a ghost. But here, as in \cite{Holdom:2021hlo}, we are working at lowest order in perturbation theory, and then the negative exclusive cross-sections that occur are a necessary part of a consistent description. We shall see more of how this works in the next section.

In any case, the fundamental property of perturbative QQG is unitarity without positivity. It is sometimes argued that it is due to unitarity that the exchange of massive degrees of freedom are not allowed to produce the signs we are seeing here. QQG shows that it is not unitarity but only positivity that needs to be relaxed. Unitarity without positivity and the resulting unusual signs that we see here are all part of a theory that produces good high energy behavior.

\section{Photon-photon scattering to two graviparticles}\label{s4}
We now consider an inelastic process, the production of graviparticles $\gamma\gamma\to{\cal G}^S{\cal G}^S$. The polarization states of the graviton are $(2e, 2o)$ and for the ghost they are $(2e, 2o, 1e, 1o, 0)$ (see appendix A of \cite{Holdom:2021hlo} for details). We start with the amplitudes for $\gamma\gamma\to gg$ from QQG, which reproduce the GR amplitudes and which are regular as $t\to0$,
\begin{align}
\label{g1}{\cal M}(\mbox{tt}2e2e)&=\frac{8\pi G}{2}\frac{t^2+u^2}{s},\\
\label{g2}{\cal M}(\mbox{tu}2e2o)&=\frac{8\pi G}{2}(u-t).
\end{align}
Each interchange t $\leftrightarrow$ u and/or $e\leftrightarrow o$ gives the same results and so we need the combination 4$\cdot$(\ref{g1})$^2$+4$\cdot$(\ref{g2})$^2$, which gives
\begin{align}
|{\cal M}|^2&=2(8\pi G)^2 \frac{t^4+u^4}{s^2}
.\end{align}
This agrees with the result in \cite{skob} for GR.

We first focus on $\gamma\gamma\to{\cal G}^S{\cal G}^S$ at asymptotically high energies. The leading contributions are given by the following representative amplitudes, omitting the factor of $8\pi G$. The amplitudes without external spin-2 graviparticles are
\begin{align}
\label{f1}{\cal M}(\mbox{tt}00)&=-m_G^2\frac{t u}{s^2},\\
\label{f2}{\cal M}(\mbox{tt}SS)&=2 m_G^2\frac{t u}{s^2},\\
\label{f3}{\cal M}(\mbox{tt}1e1e)&=\frac{1}{2} m_G^2\frac{t^2-8 t u+u^2}{s^2},\\
\label{f4}{\cal M}(\mbox{tu}1e1o)&=\frac{1}{2} m_G^2 \frac{t-u}{s}.
\end{align}
Even though $S$ is in the final state in (\ref{f2}), there is still no dependence on $m_S$. This phenomenon is also seen in \cite{Holdom:2021hlo}, where some (but not all) of the amplitudes with external $S$'s do not depend on $m_S$.

We consider amplitudes with spin-2 graviparticles in the form of the following sums,
\begin{align}
\label{f5}\sum_{\rm mass}(-1)^{n_g}\frac{p_f}{p_i}|{\cal M}(\mbox{tt}1e2e)|^2&=-2 m_G^4\frac{(t^2- \frac{1}{2}t u+u^2) (t-u)^2}{s^4},\\
\label{f7}\sum_{\rm mass}(-1)^{n_g}\frac{p_f}{p_i}|{\cal M}(\mbox{tu}1e2o)|^2&=-2 m_G^4\frac{(t^2-\frac{5}{2} t u+u^2)}{s^2},\\
\label{f6}\sum_{\rm mass}(-1)^{n_g}\frac{p_f}{p_i}|{\cal M}(\mbox{tt}2e2e)|^2&=-\frac{1}{2} m_G^4\frac{(t-u)^2 (t^2+u^2)^2}{ t u s^4},\\
\label{f8}\sum_{\rm mass}(-1)^{n_g}\frac{p_f}{p_i}|{\cal M}(\mbox{tu}2e2o)|^2&=-\frac{1}{2} m_G^4\frac{(t-u)^4}{t u s^2}.
\end{align}
For each external spin-2 graviparticle, we are summing over the two states of differing mass, that is the graviton and spin-2 ghost with masses 0 and $m_G$. This sum is performed at the level of exclusive differential cross-sections, and thus the factor of $p_f/p_i$. $(-1)^{n_g}$ is the extra minus sign that occurs when there is an odd number of ghosts. Because of these signs the sums result in cancellations that reduce the overall power of $E$ in the asymptotic behavior down to the $E^0$ behavior observed in (\ref{f5}-\ref{f8}).

The interchanges t $\leftrightarrow$ u and $e\leftrightarrow o$, as well as the interchange $1\leftrightarrow 2$ in (\ref{f5}) and (\ref{f7}), give the same results, up to a minus sign in (\ref{f4}). Thus we need the following combination to form a quantity that will yield an inclusive differential cross section for $\gamma\gamma\to{\cal G}^S{\cal G}^S$ in the high energy limit,

$\quad\quad\quad$2$\cdot$(\ref{f1})$^2$+2$\cdot$(\ref{f2})$^2$+4$\cdot$(\ref{f3})$^2$+4$\cdot$(\ref{f4})$^2$+8$\cdot$(\ref{f5})+8$\cdot$(\ref{f7})+4$\cdot$(\ref{f6})+4$\cdot$(\ref{f8}).\\The result is
\begin{align}
\frac{d\sigma^{\cal GG}}{d\Omega}&=-\frac{(8\pi G)^2 m_G^{4}}{64\pi^2 s}\frac{\left(t^{6}+\frac{11}{2} t^{5} u -7 t^{4} u^{2}-\frac{37}{2} t^{3} u^{3}-7 t^{2} u^{4}+\frac{11}{2} t \,u^{5}+u^{6}\right) }{t u s^{4}}
.\label{e8}\end{align}
This is positive for scattering into the transverse plane, $u=t=-s/2$. But there is a $1/(tu)\sim 1/\sin(\theta)^2$ pole with a negative residue, and thus the differential cross-section is not positive definite. We shall comment below on how a more inclusive process, one that also includes the elastic scattering, is positive definite.

The $1/\sin(\theta)^2$ pole in (\ref{e8}) is a result of the high energy limit $E\to\infty$. For a finite $E$, the $1/\sin(\theta)^2$ pole is regulated for $\sin(\theta)\lesssim m_G/E$. Thus for a finite $E$ the integral $\int_0^\pi (d\sigma/d\Omega) \sin(\theta)d\theta$ can be done and a finite cross-section $\sigma^{\cal GG}(s)$ for $\gamma\gamma\to{\cal G}^S{\cal G}^S$ is obtained. We do this by using the full amplitudes, not just their high energy limit.\footnote{Thus far we have calculated this for the tt photon polarizations.} The optical theorem then implies an imaginary contribution to the $\gamma\gamma$ forward scattering amplitude that we were considering in the last section, in the form of a branch cut along the real $s$ axis. Because of an intermediate state with two massless gravitons, this branch cut extends to $s=0$, as does the corresponding branch cut for negative $s$.

The cross-section $\sigma^{\cal GG}(s)$ is positive up to the $s=m_G^2$ threshold for producing a massive ghost and a soft graviton. Above this threshold the cross-section is negative, and it remains negative above a second threshold at $s=(2m_G)^2$. This cross-section displays two interesting features. One is a $s\sigma^{\cal GG}(s)\sim-\log(s)$ behavior in the high energy limit. The log is related to the fading regulation of the angular divergence as we mentioned. The other feature occurs when approaching the $s=m_G^2$ threshold from above, and it is a $1/\omega$ divergence where $\omega$ is the graviton energy. This is a standard IR divergence due to a soft graviton emission accompanying the production of the massive ghost, and it means that a one-loop correction should also be considered. It is not considered here.

Above the $s=m_G^2$ threshold the imaginary contribution to the $\gamma\gamma$ forward scattering amplitude is due to intermediate states involving both massless and massive graviparticles. This imaginary contribution can be attributed to the UV completion, and like the tree level contribution, it is once again negative. This contribution happens to satisfy the Froissart bound.

The inelastic scattering $\gamma\gamma\to{\cal G}^S{\cal G}^S$ is accompanied by the elastic scattering $\gamma\gamma\to\gamma\gamma$ as discussed in the last section. The elastic differential cross-section is positive definite and it completely dominates the inelastic differential cross-section in the forward and backward regions where the former has a singular limit. It also dominates for all other values of the scattering angle. The inclusive differential cross-section for photon-photon scattering in the high energy limit includes both elastic and inelastic scattering, and this is positive definite.

If we wish to translate this more inclusive result into a contribution to the forward scattering amplitude, we need to do the angular integration to get the inclusive cross section. But this is divergent without some way of regulating the angular divergence in the elastic scattering. This is just another consequence of the $t$-channel pole arising in $\gamma\gamma\to\gamma\gamma$ scattering due to the massless graviton, and it here obstructs the determination of the full imaginary contribution to the forward scattering amplitude. It appears that one should proceed as we did at tree-level, that is to calculate the 1-loop amplitude and then investigate the subtraction ${\cal M}_{\rm 1-loop}(s,t)-{\cal M}_{\rm 1-loop}^{GR}(s,t)$ to see whether this allows a $t\to0$ limit. But this is far beyond the present work.

\section{Conclusion}

We began this paper by giving the full differential cross section for $\gamma\gamma\to\gamma\gamma$ induced by a gravity QFT at tree-level. Its good behavior at high energies is clearly linked to the moderating effect of the ghost propagator, due to its minus sign relative to the graviton propagator. The renormalizability of the theory relies on the same effect. And then when we studied the Taylor expansion of the amplitudes, the negative definite expansion coefficients are also clearly related to the same minus sign. All of this is a consequence of how the theory enjoys unitarity without positivity.

The lack of positivity is consistent with unitarity, analyticity and locality, but there is some violation of causality \cite{Donoghue:2019ecz,Donoghue:2020mdd}. We have already mentioned that acausal behavior shows up in the dressed ghost propagator, and the timescale over which this acausality occurs is of order $1/m_G$. The negative correction $-4\pi G s^2/m_G^2$ to the forward scattering of photons that we have found here is also an indication of acausal behavior \cite{Adams:2006sv,Cheung:2014ega,Camanho:2014apa}. But if $m_G\sim m_{\rm Pl}$, which may be expected anyway, the effect is extremely suppressed. It may then be both intrinsically unobservable \cite{deRham:2020zyh} and allowed by standard arguments \cite{Caron-Huot:2021rmr}.

We found that the violation of the Froissart bound is a simple consequence of the ghost pole appearing in the $t$-channel. Since the amplitudes are known we are able to deal not only with the massive poles, but also with contributions from asymptotically large $|s|$ for fixed $t$. We then find cancellations among effects singular in $1/t$, which allows us to take the $t\to0$ limit. The result is an effective forward scattering amplitude that can be said to be due to the UV completion. In particular it includes negative imaginary contributions at the massive poles.

The lack of positivity showed up in a more nontrivial way for the inelastic production of graviparticles, $\gamma\gamma\to{\cal G}^S{\cal G}^S$. Exclusive differential cross sections are not positive definite, resulting in cancellations when summing over the two masses of graviparticles for fixed spin-2 polarizations. This yields good high energy behavior of the differential cross sections that are at least partially inclusive, as in (\ref{f5}-\ref{f8}). We were also able to obtain the cross section for $\gamma\gamma\to{\cal G}^S{\cal G}^S$, valid at all energies. It is negative above the ghost threshold, and this implies further negative imaginary contributions to the $\gamma\gamma$ forward scattering amplitude.

A positive definite differential cross section for photon-photon scattering is obtained when the inelastic and elastic ones, those for $\gamma\gamma\to{\cal G}^S{\cal G}^S$ and $\gamma\gamma\to\gamma\gamma$, are combined. This can be compared to what happens in graviparticle-graviparticle scattering where there are many inelastic and elastic processes to consider \cite{Holdom:2021hlo}. In that case only a few of the inelastic differential cross sections are negative and these few are relatively small in magnitude. Thus the more inclusive processes are very easily positive definite.

The more inclusive processes are relevant for a parton-shower-like description of ultra-Planckian scattering, as described in \cite{Holdom:2021hlo}. Hard scatterings occur via colliding partons in initial-state parton showers, and the scattered partons then produce final-state parton showers. When the colliding partons are photons then the differential cross-sections we have given here could be used, in particular to be convoluted with the photon distribution functions of the initial-state showers. The photon-photon initial state for the hard process is just one of many that occur in the interaction of the two initial-state parton showers. All possible hard processes are involved in the description of ultra-Planckian scattering.

\appendix
\section{Vertices}
For the $\gamma\gamma\to\gamma\gamma$ process we need the vertex with two on-shell photons with polarizations $e_1$ and $e_2$, and one off-shell graviparticle. It is proportional to
\begin{align}
&e_1\cdot e_2\, p_1\cdot p_2 \eta^{\mu\nu}- e_1\cdot p_2\, e_2\cdot p_1 \eta^{\mu\nu}-p_1\cdot p_2\, (e_1^\nu e_2^\mu+ e_1^\mu e_2^\nu) 
  \nonumber\\ & - e_1\cdot e_2\,( p_1^\mu p_2^\nu + p_1^\nu p_2^\mu)
  + e_1\cdot p_2\,( e_2^\nu p_1^\mu +  e_2^\mu p_1^\nu )+ 
 e_2\cdot p_1\,( e_1^\nu  p_2^\mu + e_1^\mu  p_2^\nu)
 .\end{align}
For the $\gamma\gamma\to{\cal G}^S{\cal G}^S$ process we need the vertex with one on-shell photon, one on-shell graviparticle and one off-shell photon. It is proportional to
\begin{align}
&e^\sigma E_{\alpha\beta} {p_1}^\alpha {p_1}^\beta - e^\alpha E_{\alpha\beta} {p_1}^\beta {p_1}^\sigma - e^\sigma E^\beta_\beta {p_1}^\alpha {p_2}_\alpha/2 - e^\alpha E^\sigma_\beta {p_1}^\beta {p_2}_\alpha + 
e^\alpha E^\beta_\beta {p_1}^\sigma {p_2}_\alpha/2 + e^\alpha E^\sigma_\alpha {p_1}^\beta {p_2}_\beta
\end{align}
Notice that $E_{\mu\nu}$ represents the polarization of the any of the graviparticles, $g$, $G$, or $S$, and so we cannot assume that $E_\mu^\mu=0$. We also need the vertex for two on-shell photons and two on-shell gravitons, proportional to
\begin{align}
&{e_1}^\alpha {{e_2}}^\beta {E_1}^{\mu\nu} {E_2}_{\mu\nu} {p_1}_\beta {p_2}_\alpha - 
{e_1}^\alpha {e_2}^\beta {E_1}^\mu_\mu {E_2}^\nu_\nu {p_1}_\beta {p_2}_\alpha/2 +
{e_1}^\alpha {e_2}^\beta {E_1}^\nu_\nu {E_2}_{\beta\mu} {p_1}^\mu {p_2}_\alpha -
2 {e_1}^\alpha {e_2}^\beta {E_1}_\mu^\nu {E_2}_{\beta\nu} {p_1}^\mu {p_2}_\alpha\nonumber\\& - 
2 {e_1}^\alpha {e_2}^\beta {E_1}_\beta^\nu {E_2}_{\mu\nu} {p_1}^\mu {p_2}_\alpha +
{e_1}^\alpha {e_2}^\beta {E_1}_{\beta\mu} {E_2}^\nu_\nu {p_1}^\mu {p_2}_\alpha -
{e_1}^\alpha {e_2}_\alpha {E_1}^{\mu\nu} {E_2}_{\mu\nu} {p_1}^\beta {p_2}_\beta + 
{e_1}^\alpha {e_2}_\alpha {E_1}^\mu_\mu {E_2}^\nu_\nu {p_1}^\beta {p_2}_\beta/2\nonumber\\& -
{e_1}^\alpha {e_2}^\beta {E_1}^\nu_\nu {E_2}_{\alpha\beta} {p_1}^\mu {p_2}_\mu +
2 {e_1}^\alpha {e_2}^\beta {E_1}_\beta^\nu {E_2}_{\alpha\nu} {p_1}^\mu {p_2}_\mu + 
2 {e_1}^\alpha {e_2}^\beta {E_1}_\alpha^\nu {E_2}_{\beta\nu} {p_1}^\mu {p_2}_\mu -
{e_1}^\alpha {e_2}^\beta {E_1}_{\alpha\beta} {E_2}^\nu_\nu {p_1}^\mu {p_2}_\mu\nonumber\\& +
{e_1}^\alpha {e_2}^\beta {E_1}^\nu_\nu {E_2}_{\alpha\mu} {p_1}_\beta {p_2}^\mu - 
2 {e_1}^\alpha {e_2}^\beta {E_1}_\mu^\nu {E_2}_{\alpha\nu} {p_1}_\beta {p_2}^\mu -
2 {e_1}^\alpha {e_2}^\beta {E_1}_\alpha^\nu {E_2}_{\mu\nu} {p_1}_\beta {p_2}^\mu +
{e_1}^\alpha {e_2}^\beta {E_1}_{\alpha\mu} {E_2}^\nu_\nu {p_1}_\beta {p_2}^\mu\nonumber\\& - 
{e_1}^\alpha {e_2}_\alpha {E_1}^\nu_\nu {E_2}_{\beta\mu} {p_1}^\beta {p_2}^\mu +
2 {e_1}^\alpha {e_2}_\alpha {E_1}_\mu^\nu {E_2}_{\beta\nu} {p_1}^\beta {p_2}^\mu +
2 {e_1}^\alpha {e_2}_\alpha {E_1}_\beta^\nu {E_2}_{\mu\nu} {p_1}^\beta {p_2}^\mu - 
{e_1}^\alpha {e_2}_\alpha {E_1}_{\beta\mu} {E_2}^\nu_\nu {p_1}^\beta {p_2}^\mu\nonumber\\& +
2 {e_1}^\alpha {e_2}^\beta {E_1}_{\mu\nu} {E_2}_{\alpha\beta} {p_1}^\mu {p_2}^\nu -
2 {e_1}^\alpha {e_2}^\beta {E_1}_{\beta\mu} {E_2}_{\alpha\nu} {p_1}^\mu {p_2}^\nu - 
2 {e_1}^\alpha {e_2}^\beta {E_1}_{\alpha\nu} {E_2}_{\beta\mu} {p_1}^\mu {p_2}^\nu +
2 {e_1}^\alpha {e_2}^\beta {E_1}_{\alpha\beta} {E_2}_{\mu\nu} {p_1}^\mu {p_2}^\nu
\end{align}
And then there is the $s$-channel diagram with a three graviton vertex. This vertex receives contributions from the various terms in the action (\ref{e17}), with the $R$, $R^2$ and $R_{\mu\nu}R^{\mu\nu}$ terms contributing 43, 79 and 228 terms respectively. We thus choose not to display the result. 

\acknowledgments
I am grateful to a referee for helping to improve this paper.


\begin{thebibliography}{99}

\bibitem{stelle}
 K.~S.~Stelle,    ``Renormalization   of   Higher   Derivative   Quantum   Gravity,''    Phys.~Rev.~D \textbf{16},    953   (1977).
 
\bibitem{fradkin1}
 E.~S.~Fradkin and A.~A.~Tseytlin, ``Renormalizable Asymptotically Free Quantum Theory of Gravity,'' Phys.~Lett.~ \textbf{104}B, 377 (1981). 

\bibitem{fradkin2}
 E.~S.~Fradkin and A.~A.~Tseytlin, ``Renormalizable asymptotically free quantum theory of gravity,'' Nucl.~Phys.~B \textbf{201}, 469 (1982).

\bibitem{Salvio:2014soa}
A.~Salvio and A.~Strumia,
``Agravity,''
JHEP \textbf{06}, 080 (2014)
[arXiv:1403.4226 [hep-ph]].

\bibitem{Holdom:2021hlo}
B.~Holdom,
``Ultra-Planckian scattering from a QFT for gravity,''
Phys. Rev. D \textbf{105}, 046008 (2022)
[arXiv:2107.01727 [hep-th]].

\bibitem{Dona:2015tra}
P.~Don\`a, S.~Giaccari, L.~Modesto, L.~Rachwal and Y.~Zhu,
``Scattering amplitudes in super-renormalizable gravity,''
JHEP \textbf{08}, 038 (2015)
[arXiv:1506.04589 [hep-th]].

\bibitem{Adams:2006sv}
A.~Adams, N.~Arkani-Hamed, S.~Dubovsky, A.~Nicolis and R.~Rattazzi,
``Causality, analyticity and an IR obstruction to UV completion,''
JHEP \textbf{10}, 014 (2006)
[arXiv:hep-th/0602178 [hep-th]].

\bibitem{Cheung:2014ega}
C.~Cheung and G.~N.~Remmen,
``Infrared Consistency and the Weak Gravity Conjecture,''
JHEP \textbf{12}, 087 (2014)
[arXiv:1407.7865 [hep-th]].

\bibitem{Camanho:2014apa}
X.~O.~Camanho, J.~D.~Edelstein, J.~Maldacena and A.~Zhiboedov,
``Causality Constraints on Corrections to the Graviton Three-Point Coupling,''
JHEP \textbf{02}, 020 (2016)
[arXiv:1407.5597 [hep-th]].

\bibitem{Alberte:2020bdz}
L.~Alberte, C.~de Rham, S.~Jaitly and A.~J.~Tolley,
``QED positivity bounds,''
Phys. Rev. D \textbf{103}, 125020 (2021)
[arXiv:2012.05798 [hep-th]].

\bibitem{Caron-Huot:2021rmr}
S.~Caron-Huot, D.~Mazac, L.~Rastelli and D.~Simmons-Duffin,
``Sharp Boundaries for the Swampland,''
J. High Energ. Phys. 2021, 110 (2021)
[arXiv:2102.08951 [hep-th]].

\bibitem{Brodin:2006wa}
G.~Brodin, D.~Eriksson and M.~Marklund,
``Graviton mediated photon-photon scattering in general relativity,''
Phys. Rev. D \textbf{74}, 124028 (2006)
[arXiv:gr-qc/0610025 [gr-qc]].

\bibitem{Abe:2017abx}
Y.~Abe, T.~Inami, K.~Izumi and T.~Kitamura,
``Matter scattering in quadratic gravity and unitarity,''
PTEP \textbf{2018}, no.3, 031E01 (2018)
[arXiv:1712.06305 [hep-th]].

\bibitem{skob}
V.~V.~Skobelev, ``Graviton-Photon Interaction'', Sov. Phys. J. \textbf{18}, 62 (1975).

\bibitem{deRham:2020zyh}
C.~de Rham and A.~J.~Tolley,
``Causality in curved spacetimes: The speed of light and gravity,''
Phys. Rev. D \textbf{102}, 084048 (2020)
[arXiv:2007.01847 [hep-th]].

\bibitem{Donoghue:2019ecz}
J.~F.~Donoghue and G.~Menezes,
``Arrow of Causality and Quantum Gravity,''
Phys. Rev. Lett. \textbf{123}, 171601 (2019)
[arXiv:1908.04170 [hep-th]].

\bibitem{Donoghue:2020mdd}
J.~F.~Donoghue and G.~Menezes,
``Quantum causality and the arrows of time and thermodynamics,''
Prog. Part. Nucl. Phys. \textbf{115}, 103812 (2020)
[arXiv:2003.09047 [quant-ph]].


\end{thebibliography}
\end{document}